# A Framework for Picture Extraction on Search Engine Improved and Meaningful Result

**Anamika Sharma[1], Sarita Sharma[2]**

[1] **Computer Science, MDU Rohtak, DAVIM**
**Faridabad, Haryana, India**
anamikab449@gmail.com
[2] **Computer Science, MDU Rohtak, DAVIM**
**Faridabad, Haryana, India**
sarita_kaushik@rediffmail.com
***http://www.ijcsi.org/articles/A-framework-for-picture-extraction-on-search-engine--improved-and-meaningful-result.php***

**Abstract**

Searching is an important tool of information gathering, if information is in the form of picture than it play a major role to take quick action and easy to memorize. This is a human tendency to retain more picture than text. The complexity and the occurrence of variety of query can give variation in result and provide the humans to learn something new or get confused. This paper presents a development of a framework that will focus on recourse identification for the user so that they can get faster access with accurate & concise results on time and analysis of the change that is evident as the scenario changes from text to picture retrieval. This paper also provides a glimpse how to get accurate picture information in advance and extended technologies searching framework. The new challenges and design techniques of picture retrieval systems are also suggested in this paper.

## 1. Introduction

The world has been dependent on the searching and is going to depend on it in the future. Searching is base of Learning and Learning is a never-ending process. Searching normally does on search engine in the form of text, images, news, maps, web sites etc. Learning will be more effective if it will be in the form of picture. Picture can be of text, graph, and images. Why only image searching? Simple answer is this because learning is more interactive and interesting in comparison with text. Human tendency to retention text is 20% as well as for images retention is 80%. On-line collections of images are growing larger and more common, and tools are needed to efficiently manage, organize, and navigate through them. Image searching is very helpful in every field and of any age group, leading fields are researching, learning and education and its users have been able to search the required image content with the help of the search engines like Goggle. But as the human race is moving towards the future, changes are taking place, even in the nature of the searching and the formats used. Content Base Image searching has shown ways to cope up with this change. It has helped to find the required information in easy way of learning, like images, graphs, pictures etc.

## 2. Image Searching

Today a number of search engines are available that give search facility for online database. Categories of these search engine includes web search engine, selection based search engine, Meta search engine, desktop search engine, web portal and vertical market web sites. Search Engines are information retrieval system designed to help to find information stored in online database. There are different ways of image searching. Some are based on simple searching of embedded annotations, metadata, textual context etc, while other complicated methods may include image classifications for searching that are based on color, texture, false color concepts etc. Sometimes for better precision of the resultant images, this type of search requires associating meaningful storage methods or semantic techniques that can be used further for all images of the database.

There are many types of search engines but this study is focused on query based image search engines. Before going to image search, first to find the need of image search. As per [1] there are different varieties of images available.

- Explosive growth of online image/video
- 5 billion images on web /31 million hours of TV program each year.

- Successful service like you Tube and Flickr
- Image/Video search exciting opportunity

### 2.1 Image Segmentation

Each image most likely contains multiple objects, or an object and a background. Therefore, extracting features globally is not appropriate. For this reason, as per [2] start by splitting each image into regions of similarity, using an image segmentation algorithm, with the intuition that each of these regions is a separate object in the image. Image segmentation is a well-studied problem in computer vision.

This segmentation algorithm partitions an image into similar regions using a graph-based approach. Each pixel is a node in the graph with undirected edges connecting its adjacent pixels in the image. Each edge has a weight encoding the similarity of the two connected pixels. The partitioning is done such that two segments are merged only if the dissimilarity between the segments is not as great as the dissimilarity inside either of the segments.

### 2.2 Image Annotations

Manual image labeling, known as manual image annotation, is practically difficult for exponentially increasing image database. As per [3], most of those images are not annotated with semantic descriptors, it might be a challenge for general users to find specific images from the Internet. Image search engines are such systems that are specially designed to help users find their intended images.

## 3. Crucial Concept In Image Searching

The biggest issue for image searching system is to incorporate versatile techniques so as to process images of diversified characteristics and categories. Many techniques for processing of low-level cues are distinguished by the characteristics of domain-images. As per [4], The performance of these techniques is challenged by various factors like image resolution, intra-image illumination variations, non homogeneity of intra-region and inter-region textures, multiple and occluded objects etc.The major difficulty is a gap between mapping of extracted features and human perceived semantics. The dimensionality of the difficulty becomes adverse because of subjectivity in the visually perceived semantics, making image content description a subjective phenomenon of human perception, characterized by human psychology, emotions, and imaginations. The image retrieval system comprises of multiple inter-dependent tasks performed by various phases. Inter-tuning of all these phases of the retrieval system is inevitable for over all good results.

## 4. Query Base Image Searching

QBIS is the primary mechanism for retrieving information from a database and consists of questions presented to the database. In query base image searching gives number of images matches with words present in query. To search for images, a user may provide query terms such as keyword, image file/link, or click on some image, and the system will return images "similar" to the query. The similarity used for search criteria could be meta tags, color distribution in images, region/shape attributes, etc. Most commercial image search engines fall into this category. On the contrary, collection-based search engines index image collections using the keywords annotated by human indexers. As per [5],Different approaches of QBIS includes text base,content base,context base,keybase and semantic base image searching.

## 5. Different Approaches of QBIS

- Text base query image searching
- Content base query image searching
- Context base query image searching
- Key based query image searching
- Semantic base query image searching

### 5.1. Text-Based Query Image Search Engines

Index images using the words associated with the images. Depending on whether the indexing is done automatically or manually, image search engines adopting this approach may be further classified into two categories: Web image search engine or collection-based search engine. Web image search engines collect images embedded in Web pages from other sites on the Internet, and index them using the text automatically derived from containing Web pages .

### 5.2 Content -Based Query Image Search

Content-based query image searching was initially proposed to overcome the difficulties encountered in keyword-based image search in 1990s. As per [6] Image meta search - search of images based on associated metadata such as keywords, text, etc. Content-based image Retrieval (CBIR) – the application of computer vision to the ***image search***.

### 5.3 Context Base Query Image Searching

In context base query , where searching query processe on the thesaurus meaning of words present in query .For example if the query is to find a “CAR” then ihe images can be toy car for children,brand product of car

company,image of any electonic circuit etc.It depend in what context user wants the result.

### 5.4 Key Based Query Image Serching

A Key base image searching is based on key words include in query required to transform knowledge of a passage into effective query strings in order to retrieve images from keyword based .As per example [7] consider the following passage in a child story book:

“I see three lemurs jumping around and screaming. The snake scares them. However, a sloth is still soundly sleeping. Around the corner, many children are watching a shark swimming swiftly.”

The subjects and objects in this passage include “snake”, “lemur” and “sloth”. During the first execution of the image retrieval process, the query string is formulated as “snake & lemur & sloth”.

In the case the response from the image archives indicates that there is no image annotated with all these terms, there is a need for a second query. Presumably, one reasonable strategy is to find images of a place where the “snake”, “lemur”, and “sloth” could all possibly appear

### 5.5 Semantic Base Query Image Search

The ideal CBIR system from a user perspective would involve what is referred to as semantic serching, where the user makes a request like "find pictures of dogs" or even "find pictures of Abraham Lincoln". This type of open-ended task is very difficult for computers to perform - pictures of chihuahuas and Great Danes look very different, and Lincoln may not always be facing the camera or in the same pose. Current CBIR systems therefore generally make use of lower-level features like texture, color, and shape, although some systems take advantage of very common higher-level features like faces. Although search engine is really a general class of programs, the term is often used to specifically describe systems like Google, Alta Vista and Excite that enable users to search for documents on the World Wide Web and USENET newsgroup.

According to [8] in systems extract visual features from images and use them to index images, such as color, texture or shape. Color histogram is one of the most widely used features. It is essentially the statistics of the color of pixels in an image.

Web images have rich metadata such as filename, URL and surrounding text for indexing and searching, different from traditional text-based approach, no manually labeling work is needed in current Web image search engines. The success of text-based image search engines has shown the power of textual information associated with Web images.

## 6. Shifting From Text To Image Searching

Searching is based today either on video image or text as per the requirement of the user. The Searching scenario has been totally changed as compared to the previous or only text based searching. In Image or video searching planning and thinking is important, especially for constantly changing and dynamic modules that are industry-linked and practice-oriented. Image searching is the next step in the evolution of video searching. However, Image searching offers a high degree of learning than the text searching. Now video searching, image searching and text searching comes under the umbrella of information searching.

## 7. Image Searching Issues

Searching through still images presents an interesting challenge to search engine developers. The way most search engines operate today is by appending text descriptions to the video clips and/or images, so that the searches are based on the text. As per [9] Video and photo searching is something that is still being developed and explored. For example, being able to search for an image of the Taj Mahal would be very challenging for developers, as this would require a query by image content. Basically, this means that the search engine or tool would have to be sophisticated enough to recognize an image of the Taj Mahal and differentiate it from all other possible images. Instead, one of the approaches for searching images is to search for distinctive features of an image. For example, the search tool could look for images with Taj Mahal’s features such as a architecture, doors or different types of work done on the walls of Taj Mahal. Another approach would be to search for distinctive colors of the known image. In this case, the search engine could look for the distinctive fading white color of the buildings.

## 8. Challenges In Image searching

This would present an interesting challenge for users and developers to establish a good interface for a searching tool of this type.As per [10], Efficiently searching video is even more complex than still images because now that search engines or tools have to be sophisticated enough to handle movement, lighting, and different camera angles. The searching of a video or film would have to be more sophisticated than to simply search a video frame by frame for the desired result. Users may also want to search for specific scenes in video or for zooming in and out. As we all know, user can use any kind of query (text, content, context, keywords, semantic) for image searching, so the need arises for the structured query content, so that it can be give the resultant images according to the type of query. Following problems can arises when the user give the query in text box of search engine:

1. No rules for writing of query for images.
2. Difficulties to identifying important words
3. Content base query systems have not any standard format of query image
4. Cannot mention in what context user want what type of image.
5. No Typing limitation but it takes only few important words.
6. Cost of spending lots of time scrolling through image searching result.
7. Speed of accessing the web.
8. Unwanted links / irrelevant data on searching.
9. Non-compatible sites of web (product searching sites) only.
10. Difficulties to narrowing down the semantic gap.

## 8.Implementation Technique Of Image Searching

Because of number of difficulties, we need to develop a framework that will focus on recourse identification for the user so that they can get faster access with accurate & concise results on time.

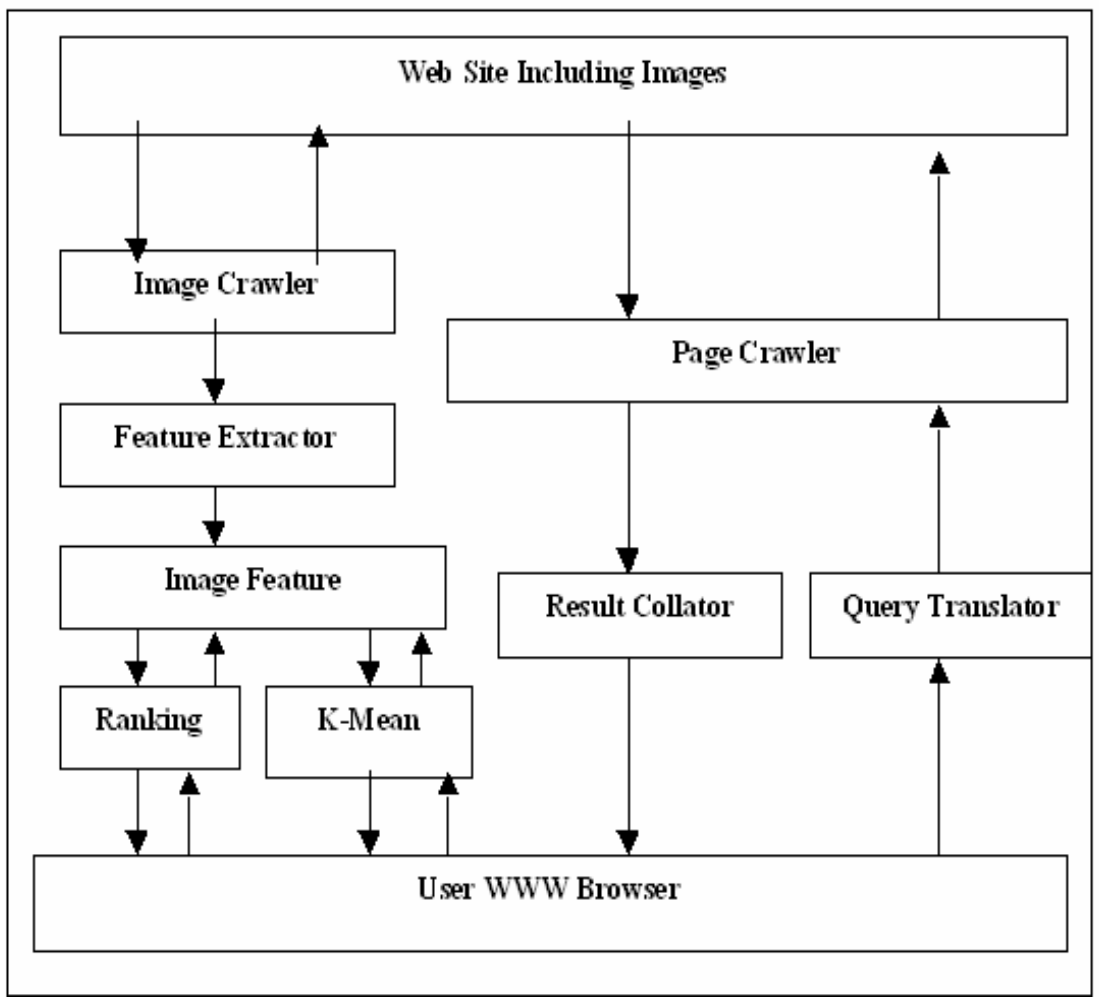

Fig.1. Framework of Picture Retrieval

Development of Framework for user to find the recourses on the bases of type of query given. Now the Proposed framework of implementation technique is to put the query on web browser. This query will translate through as

1.User input a text Query to the browser.

2.The Query Translator extracts the query from HTTP request, then translate the query into the input format for each text based image search engine.

3. The page crawler sends the query to the each search engine and collects the HTML file containing the URL of image retrieved by the search engines, then parses the HTML file to obtain the individual URL.

4. The result collator merges the result and shows the first page of retrieval image.

5. Using the URLs the image crawler retrieves the image from the Internet to construct the initial image set.

6. Feature extractor computes the image content feature vector for all images in the initial image.

7.Based on user's new request, cluster the image set using the feature vector and K-mean algorithms.

8.Based on the feedback images selected by the user, compute the distance of feature vector between the feedback image and the image in the initial image set. Re-rank images according to the distance and display the re-ranked image to the user.

## 9.Conclusion

This paper has presented an up coming wave of image searching and force for shifting from Text searching to Image searching with acceptance of new technology and directions. This includes the embedded software tools that help in online searching of content include text, image and video based on query. The related researches are also presented in this paper. With the advent of the Internet, information from all over the world is available to the people. Since there is so much information out there, people require an automated method to search through all of it. The search engines available today provide users with this ability, but primarily for text based searching. As the Internet moves further and further into being able to support multimedia, users and corporations will need to take advantage of new searching techniques. Some companies have even begun to hire "Web Specialists" to assist them in becoming aware of what searching facilities.

The idea of being able to search images, video, or audio based on the content is possible since the Internet is an electronic medium, but something that is still in development. As more and more users begin to understand the concepts behind searching these media, the need to do so will rise. However, today, most of the work being done to allow users to search this media is still in the developmental phases. The approaches described above are simply ideas and theories for ways in which this type of searching could be possible. As per [11] The idea of generating a storyboard from a video, or searching a video based on a moving sketch, or searching audio based on content and colored wave files are simply ways in which searching multimedia may become a reality.

## 10.Future Scope

Image searching provides some unique and interesting challenges for developers to come up with some sort of automated way of searching through video and/or still images. One method that is currently in development is to generate a storyboard out of a video. Storyboards typically consist "of a series of sketches showing each shot in each scene as it will be filmed, and possibly some indication of the action-taking place e.g., an arrow showing the direction of movement. A 'shot' is defined as a section of action during which the camera films continuously without interruption." As per [12] A storyboard is typically used by writers and directors while making a movie to plan the action of the shot, to review camera angles, and provide a summary of the film. Essentially, the proposed approach would be to take a finished video product and generate the storyboard based on the finish video. "In order to reverse-engineer a storyboard from the finished video sequence, it is necessary to identify three properties of each shot in the sequence. These are: (1) the start point of the shot, (2) the end point of the shot and (3) the picture that best represents the shot as a whole." As per [12] Once the storyboard has been generated, it will be easier to search for video sequences, especially in large video libraries.

Another approach to video searching is the search actual video using video cues. At Columbia University, a system called VideoQ is being developed that does this. The theory behind this system is to have the user actually draw out an animated scene as the query. "In an animated sketch, motion and temporal duration is the key attributes assigned to each object in the sketch in addition to the usual attributes such as shape, color, and texture. Using the visual palette, we sketch out a scene by drawing a collection of video objects." As per [13] The VideoQ system will then search its video library for videos that match the animated sketch. The VideoQ system is intended to be on the Internet and use various Java applets to allow the user to create these animated sketches.

**Ms. Anamika Sharma** did her Master in computer Application from Gurukul University Haridwar in 1998 ,M.Tech From Allahabad Agriculture University,M.Phil from Vinayka Mission University Tamil Nadu and pursuing Ph.D from Singhania University Rajasthan. She is having about 13 years of teaching experience of postgraduate courses. She has guided more than hundred students in their project and published number of papers in national/International journals. She is a member of Computer Society of India .Her main Area of research include Computer Graphics, Data mining, Software testing and Quality assurance and object oriented Analysis and Design. At Present she is working in DAV Institute of Management, Faridabad as Associate Professor in Computer Science Deptt.

**Ms. Sarita Sharma** did her Master in Computer Application from IGNOU, New Delhi, India, She did her M.phil (Computer Science) from Ch. Devi Lal University,Sirsa , India and is pusuing Ph.D from Singhania University,Rajasthan, India. She is presently working as Associate Professor in Deptt. Of Computer Science, DAV Institute of Management , India. She has guided more than 90 students in their Projects and has published a number of papers in National/International journals. She is a member of Computer Society of India. Her areas of interest include Software Engineering, Data Mining, Relational Databases, Computer Languages etc. She has about 15 years of teaching experience. with current employment; association with any official journals or conferences